# Perturbations to the Cosmological Expansion in a Grainy Universe

Brett Bochner

Dept. of Physics and Astronomy,

Hofstra University

*(phybdb@hofstra.edu,*

*brett_bochner@alum.mit.edu)*

# Abstract:

The matter content of the Universe is generally regarded as a perfect fluid on sufficiently large scales, for all epochs. But the recent cosmological matter distribution, consisting of an (ideally) random distribution of gravitationally collapsed structures, is more accurately described as a collection of discrete grains, than as a fluid.

It is well known that granular materials may have very different macroscopic properties than fluids; analogously, we investigate the possibility that pervasive small-scale inhomogeneities in the recent Universe may lead to perturbations of the cosmological expansion on intermediate and/or large scales.

# Research Rationale:

Two of the cornerstones of modern Cosmology have been the assumptions of *Isotropy* and *Homogeneity* for the distribution of mass, energy, and pressure in the Universe. These assumptions have been crucial for simplifying Einstein's equations, resulting in the development of the Friedmann equation; and they have been vindicated by the success of Big Bang theory, the observed homogeneity of the CMB, and the concordance which now appears to be growing in the determination of Cosmological Parameters.

Beyond such successes, however, these two assumptions are very restrictive towards deeper efforts at deciphering the evolutionary history of the Universe, tying all theoretical considerations to one very simple dynamical equation (the Friedmann equation). Fundamental instabilities in (and/or peculiarities of) Friedmann evolution are to some degree responsible for all of the well-known "esthetic" problems with standard cosmology -- e.g., the Horizon & Flatness Problems, the Cosmological Constant Problem, the Coincidence Problem, the still somewhat mysterious Acceleration of the Universe, and so on.

Various well-regarded theories exist, which attempt to deal with these problems -- e.g., Inflation theory, zero-point energy cancellations from Supersymmetry, Tracker Quintessence models, etc. -- but these solutions invariably depend upon yet-unproven theories, undiscovered particles, unknown or poorly-defined mechanisms, and the assumption that the matter content of our Universe is dominated by (at least) two exotic particle species, one being largely unobserved (Dark Matter), and the other being largely counterintuitive (Dark Energy). In response to this, we ask: would it be possible to obtain simpler solutions to esthetic Cosmological problems by permitting the Universe a *richer dynamical evolution*? In other words, could relaxing the assumptions of Isotropy and/or Homogeneity allow us to go beyond the strictures of the Friedmann equation?

In prior research, we have investigated the effects of large-scale *anisotropy* in the very early Universe. Here, we consider the idea that the nonlinear, *inhomogeneous clumping* of material in the *"recent"* Universe may alter the Friedmann expansion, locally or perhaps even globally. Such considerations have at least one immediate advantage: since the large-scale gravitational clumping of matter is a "recent" Cosmological phenomenon, it is natural to apply it to recent Cosmological effects, such as the *acceleration* of the expansion of the Universe.

# Problem considered here: *Cosmological Acceleration*

The FRW acceleration equation:

$$\ddot{R} = -\frac{4\pi G}{3} R \cdot \sum_{i}^{\text{All Species}} (\rho + 3P)_i$$

=> Acceleration ($\ddot{R} > 0$) is achieved by species with *sufficiently negative pressure*, $P < -\rho/3$.

- Standard Hypothesis:

=> Universe is filled with a "repulsive" force/energy density/ particle species, "Dark Energy", accelerating its expansion.

=> "D.E." fills the gap between total mass-energy density implied from clustering ($.3 < \Omega_m < .4$) & CMB ($\Omega_{\text{Tot}} \cong 1$) studies.

# Dark Energy & Large-Scale Structure: Does D.E. "*Clump*"?

- *Crucial* Property of D.E.: It *must not* clump, or would be *ruled out* by large-scale galaxy clustering studies.

  => Standard (*erroneous*) assumption: *"Repulsive"* action of *Negative Pressure* pushes D.E. particles apart (to avoid clumping), the same way it "pushes" the Universe to accelerate.

- Error seen from simple application of 1$^{st}$ Law of Thermodynamics:

$$P = -(\partial E/\partial V), \quad P < 0 \implies (\partial E/\partial V) > 0 \; !$$

  => It *takes work* to expand a volume of D.E. material ... "Negative Pressure" comes from substance that *attracts*, not *repels*!

  => Result is counterintuitive: what attracts *locally*, repels *globally*.
  (Due to "backwards" sign, $\ddot{R} \propto -P$, in acceleration equation.)

# Problem w/Dark Energy Scenarios: *Clumping* of D.E. Particles

- <u>Big Problem</u>: Not only would D.E. particles clump together on sub-horizon scales, but would do so *relativistically*, since $|P| \approx \rho$.

  => D.E. may be the *most strongly clumping* material in the Universe!

  => Problem *not solved* by postulating some additional force (e.g., Degeneracy Pressure) to keep D.E. particles apart; that just introduces a source of *Positive Pressure*, <u>negating</u> D.E.'s Neg. Press. & Acceleration.

  => Universe *expends work* creating D.E. particles in expansion; energetically favorable for them to clump, to *reduce volume* of space containing D.E. fields (i.e., facilitate particle annihilation).

- Problem not necessarily universal for *all* D.E. scenarios... e.g.:

  => Pure "Cosm. Constant" energy ($\Omega_\Lambda$), being *immobile*, cannot clump.

  => "Solid" D.E. [e.g., M. Bucher, D. Spergel, 19$^{th}$ Tex. Symp.] also can't clump.

# A Virtue of Necessity: Could *clumped matter* cause the Acceleration?

- Any particle/mechanism which *extracts energy* from (or "fights") the Cosmological expansion, would exhibit "Negative Pressure".

- Normal *Gravitational Attraction* is a form of "Negative Pressure"!

  => Nonlinear *gravitational collapse* of matter (i.e., structure formation) on intermediate to large scales is cosmologically "recent" -- and *acceleration* is "recent" -- Could the *former* be causing the *latter*?

  => The recent onset of acceleration would be "normal"; no need to invoke any "trigger" or fine-tuning mechanism (i.e., Tracker Quintessence).

  => Being recent, *acceleration* is not observed (or *observable*) out to high-Z; it may therefore be a "local" phenomenon, not "global", like other cosmological phenomena studied at the CMB level.

  *(Is it possible that only "local bubbles" of universe, encompassing large clusters of matter, experience an accelerating expansion?)*

# Difficulties with this "Clumped Matter" Hypothesis:

- If acceleration comes from $\Omega_m$, and not "$\Omega_{D.E.}$", than *what could* $\{1 - \Omega_m \cong .6 - .7\}$ *be?* (& how could *that* material avoid clustering?)

  => Neglect this issue for now; our purpose here is to *gain perspective* by taking a step *away* from "concordance", at least temporarily.

- Could $\Omega_m$ generate *enough* Neg. Pressure to cause acceleration?

  => Hard to imagine... $|P| \approx \rho$ is needed (*relativistic grav. collapse!*).

- Maybe use relativistic, powerfully attractive objects... *Black Holes!*

  => Black Holes seem increasingly to be *very common* in the Universe... yet, it seems likely that $\Omega_{B.H.} \ll 1$, requiring $|P_{B.H.}| \gg \rho_{B.H.}$ -- a violation of the *Dominant Energy Condition* (DEC) -- to achieve the condition of $|P_{B.H.}| > \rho_{Tot}/3$ needed for acceleration!

  => It is *conceivable* that BH's may violate the DEC, because: (i) BH's are "exotic material"; (ii) No "motion" here => *causality* is *not threatened*.

# Hypothesis: Could *Black Holes* power the Cosmological Acceleration?

- BH's could conceivably *extract energy* (i.e., $P < 0$) from the expanding Universe, if they are "stretched" along with the expansion…

    => Innumerable BH's embedded in the FRW Universe would contribute.

    => Resulting acceleration could be "patchy", not smooth or uniform.

    => Resulting acceleration would not necessarily be "exponential" or *eternal*, but would be *slowed* by *dilution* of BH number density in a growing Universe (…or, contrarily, *sped up* during "*rapid*" *BH formation*).

- To check idea, would like to solve "test case": *1 BH in FRW Universe.*

=> Problem (in progress) is *very difficult*, and literature on problem is deep; no direct solution for us yet. Consider a *"general"* possibility here:

  - BH *expanding proportionally* with the Universe (i.e., maintaining constant "coordinate volume"): $R_{\text{Schwarzchild}}(t) \propto R_{\text{FRW}}(t)$

# Toy Model: Black Hole embedded in a FRW Expanding Universe

- Effect of Expansion on BH Size (& "Mass"):

$$R_{\text{Schwarzchild}} = 2GM_{\text{BH}}/c^2 \implies R_{\text{Sch.}} \propto M_{\text{BH}}$$

=> B.H. expansion equiv. to "mass-energy increase" (taken from Universe)?

- 2 "obvious" guesses to make for how BH's *physical size* scales w/ $R_{\text{FRW}}$:

  (i) $R_{\text{Sch.}} \propto Constant$ -- i.e., BH is *unaffected* by the FRW Cosmological Expansion... and the BH *has no effect* <u>on</u> the Expansion.

  (ii) $R_{\text{Sch.}} \propto R_{\text{FRW}}$ -- i.e., Cosm. Expansion *stretches* BH... larger $R_{\text{Sch}}$ may be equivalent to larger $M_{\text{BH}}$... mass-energy difference is *taken* from FRW Expansion... => BH acts as *Neg. Pressure source!*

- So... *does* the FRW Expansion stretch out small, *"local"* objects?

  => *Many* scientists (including Einstein) have argued this issue, with no definitive answer... we'll consider the *more interesting* possibility, (ii), here...

# Simple Estimate: Deriving the "Effective Negative Pressure" of a BH

- "Effective Energy" of 1 BH:

$$r_{Sch.} = 2GM_{BH}/c^2 \implies E_{BH} \equiv M_{BH}c^2 = r_{Sch.}c^4/(2G)$$

- "Eff. Pressure" of BH in Co-moving Cosm. Volume, $V = (4/3)\pi R_{FRW}^3$:

$$P_{BH} = -\partial E_{BH}/\partial V = (-\partial E_{BH}/\partial R_{FRW}) \div (4\pi R_{FRW}^2)$$

- Assuming BH *expands proportionally* with the Universe:

$$r_{Sch.} = r_0 \cdot (R_{FRW}(t)/R_0) \implies P_{BH} = (-c^4/8\pi G)(r_0/R_0)(R_{FRW}^{-2})$$

- "Eff. Energy Density" of BH, using $\rho_{BH} \equiv E_{BH}/V$:

$$\rho_{BH} = (3c^4/8\pi G)(r_0/R_0)(R_{FRW}^{-2}) \implies (P_{BH}/\rho_{BH}) \equiv w = -1/3$$

==> <u>*Not Enough Negative Pressure to cause Acceleration*</u>

(recall p. 5); just enough for "Coasting" (i.e., $\ddot{R} = 0$) !

## *Do* BH's cause acceleration?: Another (related) estimate of BH *Neg. Press.*

- *Total* BH Mass Density in some co-moving volume, V, might be:

$$\rho_{BHs, Tot} = N_{BH} \cdot M_{each\ BH} / V$$

Now using certain simplifying assumptions/relations:

- $R_{Sch.}(t) \propto R_{FRW}(t) \implies M_{each\ BH}(t) \propto R_{Sch.}(t) \propto R_{FRW}(t)$

- $V \propto R_{FRW}(t)^3$

- Assuming (*for now*) the *(probably incorrect)* simplification of <u>no new BH creation</u>: $N_{BH} \propto Constant \propto R_{FRW}(t)^0$

$$\implies \rho_{BHs, Tot} = R_{FRW}(t)^{-2} \ !$$

- Knowing that a species w/ $P = w \cdot \rho$ evolves like $\rho \propto R_{FRW}^{-3(1+w)}$,

$\implies w_{BHs} = -1/3$ , $P_{BHs, eff.} = -\rho_{BHs}/3$ ... <u>*Not enough for Acceleration!*</u>

# Hope for the BH Hypothesis: Could BH's *more powerfully* induce Acceleration?

These *simple estimates* indicate that BH's are <u>*insufficient*</u> to cause Cosmological Acceleration. Is there any way to avoid this conclusion?

- Discard assumption of *constant co-moving BH number density*...

  => Would <u>*rapid production*</u> of *new* BH's in the late-evolving Universe lead to "more effective" negative pressure...? (i.e., $|P_{B.H.}| \gg \rho_{B.H.}/3$)

- Might BH's be *far more common* in the Universe than we even yet realize, exerting much more Neg. Press.? *(Seems far-fetched, but...)*

  => Could $\Omega_{B.H.} \approx \Omega_m \approx \Omega_{D.E.}$, rather than $\Omega_{B.H.} \approx$ (fraction of %) $\times \Omega_m$ ...?

- Other Possibilities...

  => Stronger effects for *spinning* (Kerr) BH's? ; Stronger "Neg. Press." effect for BH's w/ *extreme masses* (mini- or supermassive BH's)? ; etc...

- Maybe "toy model" incorrect: *An exact, quantitative solution for the evolution of a BH embedded in a FRW Universe would be useful!*

# Summary and Conclusions:

- By discarding assumptions of *isotropy* and *homogeneity* under certain circumstances, we seek to go beyond the Friedmann Equation, to find *new* explanations for the "esthetic" problems in modern Cosmology.

- We note that *Negative Pressure* implies attractive forces, <u>not</u> repulsive forces; this is a problem for *Dark Energy*, since it implies that D.E. will *strongly cluster* if it is composed of *mobile particles*.

- Correspondingly, we find that *strongly clumped matter* will possess *Negative Pressure*; we consider that matter with relativistically attractive gravitational forces -- i.e., *Black Holes* -- might conceivably provide enough Negative Pressure to *Accelerate* the expansion of the Universe.

- Lacking a full solution to the evolution of a BH in a FRW Universe, we make qualitative estimates, assuming that the BH *stretches proportionally* with the Universe; we find that BH's exert *significant* Neg. Press. in that case, but *not enough* for acceleration, under these *simplifying assumptions*.

=> In any case, the unknown effect of <u>*pervasive inhomogeneities*</u> on the overall *"FRW-like" expansion* represents an interesting, open question!

# Comments or Questions…?